\documentclass[%
reprint,
showkeys,
superscriptaddress,
preprintnumbers,
nofootinbib,
amsmath,amssymb,
aps,
prd,
floatfix,
]{revtex4-2}

\usepackage{afterpage}
\usepackage{subfigure}
\usepackage{rotating} 
\usepackage{booktabs}
\usepackage{graphicx}
\usepackage{dcolumn}
\usepackage{bm}
\usepackage{amsmath}
\usepackage{natbib}
\usepackage{hyperref}

\begin{document}

\preprint{Published: Q.~Liu, H.~Song, B.-Q.~Ma, \href{https://doi.org/10.3847/2515-5172/ad85e7}{Res. Notes AAS 8 (2024) 263}}

\title{Direct evidence for preburst stage of gamma-ray burst from GRB 221009A data}

\author{Qing Liu}

\author{Hanlin Song}

\affiliation{School of Physics, Peking University, Beijing 100871, China}

\author{Bo-Qiang Ma}\email{mabq@pku.edu.cn}
\thanks{corresponding author.}

\affiliation{School of Physics, Peking University, Beijing 100871, China}
		
\affiliation{Center for High Energy Physics, Peking University, Beijing 100871, China}

\begin{abstract}
Previous research on Lorentz invariance violation in photons from gamma-ray bursts (GRBs) suggested a scenario where multi-GeV photons could be emitted before lower-energy photons at the GRB source frame. This implies the existence of a new preburst phase in addition to the traditionally identified prompt and afterglow stages observed in earlier studies. In this study, we present direct evidence for this novel preburst phase in gamma-ray bursts based on recent observations of GRB 221009A. Our analysis leverages data from the Fermi Gamma-ray Burst Monitor (GBM) and Large Area Telescope (LAT) detectors of the Fermi Gamma-ray Space Telescope (FGST), as well as data from the KM2A detector of the Large High Altitude Air-shower Observatory (LHAASO).

\end{abstract}
\keywords{gamma-ray burst; preburst stage;  high-energy photon; Lorentz invariance violation; GRB 221009A}

\maketitle





Gamma-ray bursts (GRBs), recognized as the most energetic events following the Big Bang, have captivated researchers for over five decades. Initially discovered in the late 1960s by the Vela satellites, the observation of intense bursts of gamma rays 
prompted further investigations into the enigmatic nature and origins of these cosmic phenomena. Over the years, extensive research and technological advancements have significantly enhanced our understanding of GRBs, dividing their evolution into two distinct stages: the prompt and afterglow phases, as traditionally identified from earlier observations.

Early satellite missions focused on exploring GRBs within the keV and MeV gamma-ray bands. The advent of the Fermi Gamma-ray Space Telescope (FGST)~\cite{Fermi-LAT:2009ihh,Meegan:2009qu} revolutionized GRB studies by enabling the observation of multi-GeV photons, while the Large High Altitude Air-shower Observatory (LHAASO)~\cite{LHAASO:2019qtb} has extended the exploration to multi-TeV GRB photons. These observatories mark crucial milestones in investigating different energy ranges of GRB photons, with FGST targeting multi-GeV photons and LHAASO focusing on multi-TeV photons. This presents an opportune moment to delve into the characteristics and features of these multi-GeV and multi-TeV photon emissions associated with GRBs.

The abundance of GRB observations, coupled with the rise in multi-GeV photon events detected by FGST, has sparked active research into Lorentz invariance violation of photons, yielding significant progress across various theoretical frameworks (for recent reviews, see, e.g. \cite{He:2022gyk,AlvesBatista:2023wqm}). Analysis of multi-GeV photon events from several GRBs by FGST~\cite{Xu:2016zxi,Xu:2016zsa,plb138951} has revealed a scenario suggesting the early emission of multi-GeV photons preceding lower-energy photons at the GRB source frame. However, this scenario is met with challenges, as direct observations indicate that these multi-GeV photons are recorded after the trigger time of GRBs~\cite{Fermi-LAT:2009owx}. While some signals hint at the observation of lower-energy GeV-scale photons before the prompt stage~\cite{Zhu:2021pml,Chen:2019avc,Zhu:2021wtw}, concrete evidence for the predicted preburst stage of multi-GeV photons in GRBs remains elusive.

Against this backdrop, a closer examination is warranted to ascertain signals or evidence for the preburst stage of multi-GeV photons observed by FGST~\cite{Lesage:2023vvj} and of multi-TeV photons observed by LHAASO~\cite{LHAASO:2023lkv}, focusing on the recent GRB 221009A, which stands out as the most potent gamma-ray burst~\cite{Lesage:2023vvj,LHAASO:2023kyg,LHAASO:2023lkv}, characterized by the detection of the highest-energy photons~\cite{LHAASO:2023lkv}. The unprecedented energy release and remarkable photon energy levels emitted during GRB 221009A offer valuable insights into the extreme astrophysical processes underlying these cosmic phenomena, shedding light on the mechanisms driving such intense bursts of gamma-ray radiation. 
In this note, we consider direct evidence which emerges when examining
the multi-GeV photon events observed by the LAT detector of FGST~\cite{Lesage:2023vvj} and the multi-TeV photon events observed by the KM2A detector of LHAASO~\cite{LHAASO:2023lkv}.

FGST is a NASA space observatory designed to study gamma-ray sources in the universe~\cite{Fermi-LAT:2009ihh,Meegan:2009qu} and 
consists of two main instruments: the Large Area Telescope 
(LAT)~\cite{Fermi-LAT:2009ihh} and the Gamma-ray Burst Monitor (GBM)~\cite{Meegan:2009qu}.

LAT is the primary instrument on FGST and is designed to detect gamma rays in the energy range from 20 MeV to about 300 GeV. It provides high-resolution imaging and spectroscopy of gamma-ray sources, allowing for detailed studies of gamma-ray bursts, active galactic nuclei, pulsars, and other high-energy phenomena. 
GBM is a secondary instrument on FGST that is specifically designed to detect and study GRBs. It consists of 14 scintillation detectors that cover a wide energy range from a few keV to 40 MeV. The GBM provides rapid alerts and localization of GRBs, allowing for follow-up observations by other ground- and space-based telescopes. 


\begin{figure*}[htbp]
    \centering
    \begin{minipage}{0.7\textwidth}
        \centering
        \includegraphics[width=0.8\linewidth]{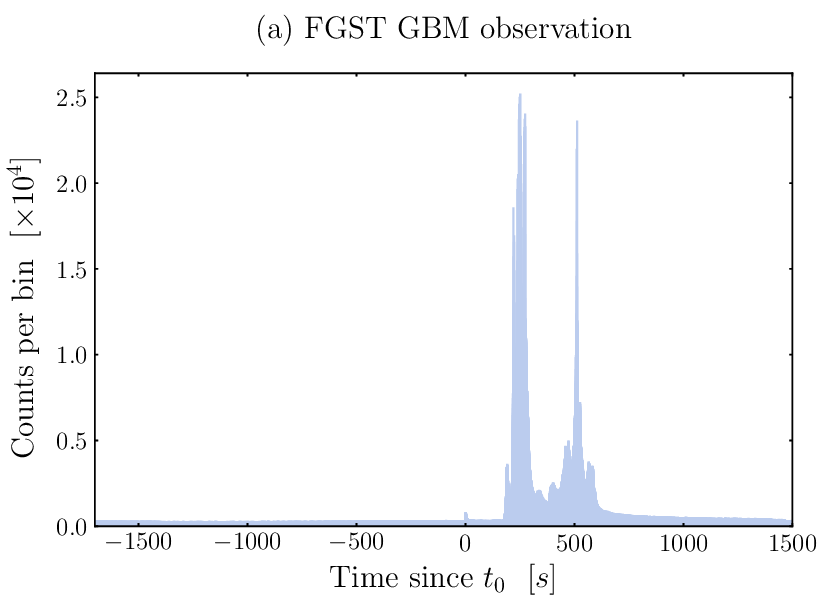}
    \end{minipage}
    
    \vspace{0.1em}

    \begin{minipage}{0.7\textwidth}
        \centering
        \includegraphics[width=0.8\linewidth]{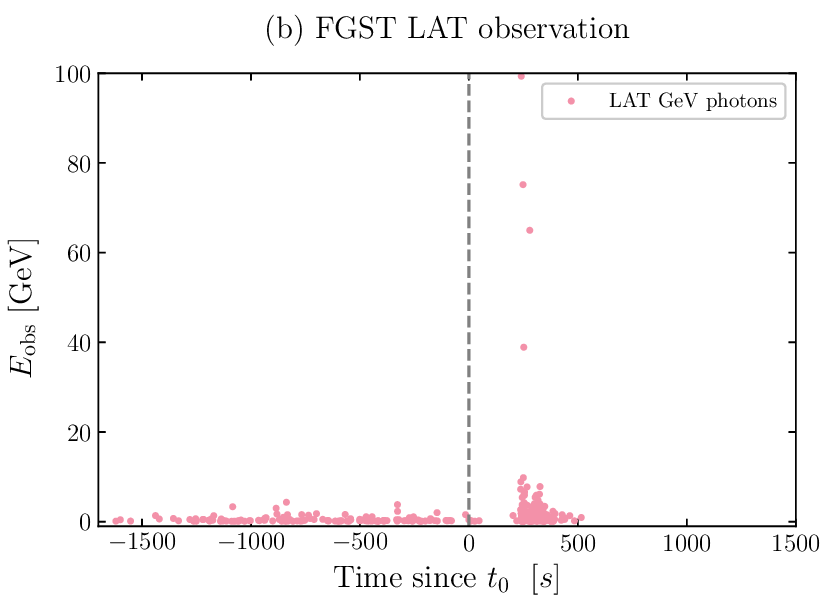}
    \end{minipage}
    
    \vspace{0.1em}

    \begin{minipage}{0.7\textwidth}
        \centering
        \includegraphics[width=0.8\linewidth]{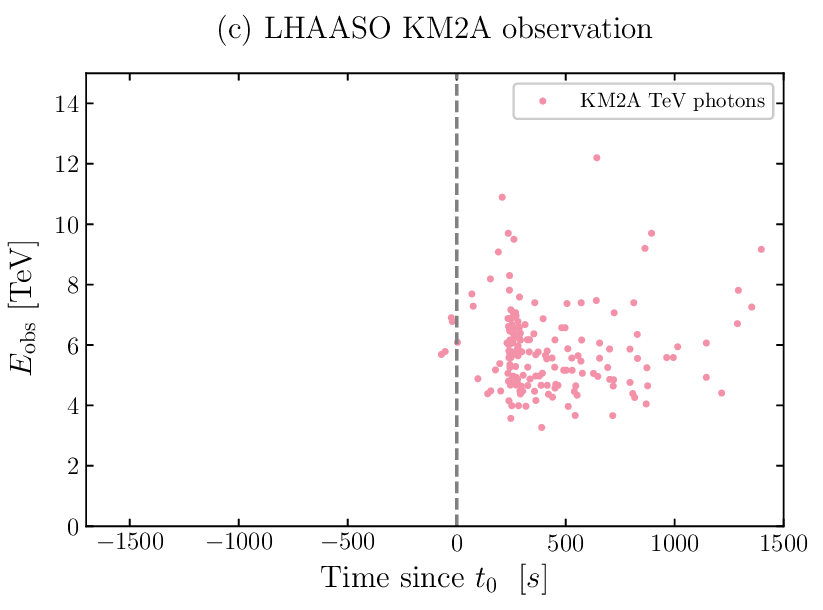}
    \end{minipage}

    \caption{Subfigure (a) shows the light curve of GRB 221009A from the observation of Fermi Gamma-ray Burst Monitor (GBM NaI-n4, NaI-n6, NaI-n7 and NaI-n8, with photon energy range of 8 to 260 keV). The time bin is chosen as 64 ms. Subfigure (b) shows the multi-GeV photons of GRB 221009A from the observation of Fermi Large Area Telescope (LAT, with photon energy range of 0.1 to 100 GeV). The vertical dashed line denotes the trigger time of Fermi Gamma-ray Burst Monitor \cite{Lesage:2023vvj}. Subfigure (c) shows the released multi-TeV photons of GRB 221009A detected by KM2A detector of LHAASO \cite{LHAASO:2023lkv}, with energies larger than 3 TeV and time period during -70 to 1400 seconds.}
    \label{fig_observations}
\end{figure*}

Through the analysis of GBM photons from GRB 221009A within the energy range of 8-260 keV, we have constructed the light curve depicted in subfigure (a) of Fig.~\ref{fig_observations}, revealing the following distinct features:
\begin{itemize}
\item 
A faint peak is observed at the trigger time (t=0) of GRB 221009A, extending from t=0 to 10 seconds.
\item 
The initial prominent peak occurs at t=190 seconds, spanning the interval from 180 to 210 seconds.
\item 
A subsequent significant peak is noted at t=220 seconds, lasting from 210 to 230 seconds.
\item 
The third and most substantial peak is observed at t=250 seconds, occurring between 230 and 260 seconds.
\item 
A fourth notable peak emerges at t=275 seconds, within the timeframe of 260 to 300 seconds.

\item 
Additionally, a significant sharp peak is evident around t=500 seconds, with some fluctuations observed during the period from t=400 to 600 seconds.

\item 
Minimal signal is detected during the timeframe of -1000 to 0 seconds, with a 
a slowly declining background
observed from 600 to 1000 seconds.
\end{itemize}

Furthermore, upon examining the LAT photon events from GRB 221009A, we search the data with equatorial coordinates being (288.266, 19.773) degrees \cite{Lesage:2023vvj, 2022GCN.32636....1V,Zhu:2022usw} 
within a 12$^{\circ}$ region of interest (ROI)~\cite{Fermi-LAT:2013oro}, 
with the energy range of 0.1 to 100 GeV over the period from -3000 to 3500 seconds relative to the trigger time. In total 371 high-energy photons are obtained, and we plot the results in  subfigure (b) of Fig.~\ref{fig_observations}, revealing the following features:
\begin{itemize}
\item 
Events distributed at a roughly constant rate
are detected from t=-1700 to 50 seconds.
\item 
There is a notable absence of events during the period from t=50 to 180 seconds.

\item 
A higher frequency of events is observed from t=190 to 500 seconds, with a sharp peak around 240 seconds.

\item 
No events are recorded during the timeframe of -3000  to -1700 seconds.

\item 
Similarly, there are no events detected from t=600 to 3500 seconds.
\end{itemize}

By comparing the findings from subfigure (a) and subfigure (b) from Fig.~\ref{fig_observations}, it becomes evident that LAT GeV-scale events are detected during the period from -1700 to 50 seconds, preceding the faint trigger signal of GBM keV+MeV photons. Moreover, LAT events are observed before the main peaks of GBM photons from t=190 seconds onwards. This direct evidence from FGST detection with LAT data supports the existence of a preburst phase of GeV-scale photons in GRB 221009A.

LHAASO is a comprehensive cosmic ray observatory located in Daocheng, Sichuan, China~\cite{LHAASO:2019qtb}. It is designed to study high-energy cosmic rays, gamma-rays, and other astrophysical phenomena. LHAASO comprises several different types of detectors, each serving a specific purpose in the observation of cosmic rays and gamma-rays, among which the KM2A detector array is designed to detect very high-energy gamma-rays in the multi-TeV energy range. It consists of a large number of imaging atmospheric Cherenkov telescopes (IACTs) distributed over a wide area, allowing for the precise measurement of the gamma-ray properties and the identification of their astrophysical sources. 




We present the distribution of multi-TeV photon events from GRB 221009A detected by the KM2A detector of LHAASO~\cite{LHAASO:2023lkv}, as illustrated in subfigure (c) of Fig.~\ref{fig_observations}. Our observations reveal the following key findings:
\begin{itemize}
\item 
A prominent peak is observed at t=240 seconds, with photon events spanning from -70 to 1400 seconds.
\item 
A noteworthy total of at least 16 multi-TeV events are identified prior to t=230 seconds.
\item 
A minimum of 10 multi-TeV events are recorded before t=180 seconds.

\item 
There are also several events before the trigger time t=0 of Fermi Gamma-ray Burst Monitor on GRB 221009A. 
\end{itemize}

In particular, a substantial number of multi-TeV photons (168 events with energies exceeding 3 TeV) are observed within the timeframe of -70 to 1400 seconds relative to the trigger time~\cite{LHAASO:2023lkv}. This dataset unequivocally showcases the detection of multiple high-energy photons preceding the emergence of keV-MeV photons observed by FGST-GBM.
These observations provide direct evidence supporting the notion that a significant emission of multi-TeV photons occurs prior to the appearance of lower-energy photons during the prompt burst phase at the GRB source, even in the absence of Lorentz invariance violation.

Thus, our findings offer robust evidence for the intriguing preburst stage associated with GRB 221009A, reinforcing the existence of a preburst phase in gamma-ray bursts as indicated by the investigation~\cite{Xu:2016zxi,Xu:2016zsa,plb138951} into Lorentz invariance violation of photons through the analysis of multi-GeV photon events from various GRBs observed by FGST. 

However, our new observation of the preburst phase 
is supported by the direct observations of GeV photons by FGST-LAT and of TeV photons by LHAASO-KM2A. This discovery serves as compelling evidence, independent of Lorentz violation considerations, and holds significance for advancing our understanding of GRBs and studying Lorentz violation in photons within these cosmic events.


In summary, 
the study of gamma-ray bursts (GRBs) 
for over five decades with
advances in observational technology and theoretical models have led to the division of GRBs into distinct prompt and afterglow phases, shedding light on the evolution of these high-energy events.
The Fermi Gamma-ray Space Telescope (FGST) and the Large High Altitude Air-shower Observatory (LHAASO) have been instrumental in observing multi-GeV and multi-TeV photons, respectively, sparking investigations into the preburst stage and Lorentz invariance violation in GRBs. Analysis of GRB 221009A has provided direct evidence for the preburst phase of multi-GeV and multi-TeV photons, supporting the existence of this phenomenon and offering valuable insights into the extreme astrophysical processes driving GRBs. This research highlights the significance of high-energy photon emissions in unraveling the complexities of GRBs and understanding their intense gamma-ray radiation bursts.



\emph{\textbf{Acknowledgements}.---}%
This work is supported by National Natural Science Foundation of China under Grants No.~12335006 and No.~12075003.

\bibliography{scibib}

\end{document}